\documentstyle[12pt]{article}
\def\beq{\begin{equation}}   \def\eeq{\end{equation}}
\def\be{\begin{equation}}   \def\ee{\end{equation}}
\def\bea{\begin{eqnarray}}   \def\eea{\end{eqnarray}}

\begin{document}

\begin{flushright}
UND-HEP-01-BIG\hspace*{.2em}08\\
hep-ph/0112155\\
\end{flushright}
\vspace{.3cm}
\begin{center} \Large 
{On A Recent Claim Concerning $\tau (D_s)/\tau (D^0)$ }\\
\end{center}
\vspace*{.3cm}
\begin{center} {\Large 
I. I. Bigi \\
\vspace{.4cm}
{\normalsize 
{\it Physics Dept.,
Univ. of Notre Dame du
Lac, Notre Dame, IN 46556, U.S.A.} }   
} \\
\vspace{.3cm}
e-mail address:\\
{\it bigi.1@nd.edu} 
\vspace*{1.4cm} 

{\Large{\bf Abstract}}\\
\end{center} 
It has been suggested that the observed 
$D_s - D^0$ lifetime difference can be `understood' as 
due to $SU(3)_{Fl}$ breaking in the phase spaces for the 
exclusive $D^0$ and $D_s$ channels `removing' the need for weak 
annihilation. It is pointed out that several of 
the statements in this argument are misleading or even 
inconsistent with a state-of-the-art description of heavy 
flavour decays based on the operator product expansion.

\vspace*{.2cm}
\vfill
\noindent
\vskip 5mm
\pagebreak 
The authors of Ref. \cite{NUSS} have argued that the 
observed difference in the lifetimes for $D^0$ and 
$D_s$ mesons, namely $\tau (D_s)/\tau (D^0) \simeq 1.17$, can be 
understood as a phase space effect {\em without} a need for what is 
usually referred to as (weak) annihilation contribution. 
Unfortunately the authors seem to be quite unaware of the 
extensive theoretical literature on this subject over the last 
decade. It is thus not surprising that they make several 
misleading or even wrong statements concerning the 
theoretical description. I will first recapitulate briefly 
the main features of the theoretical description before listing 
the specific criticism of the claims put forward in 
Ref.\cite{NUSS}.  
\section{Basics of the Heavy Quark Expansion (HQE)}
A state-of-the-art description of the weak decays of heavy 
flavour hadrons $H_Q$ has to be based on the operator product 
expansion (OPE), which allows to express decay widths -- 
semileptonic, radiative or nonleptonic ones -- through an expansion 
in powers of $1/m_Q$ with $m_Q$ denoting the heavy flavour 
{\em quark} mass \cite{HQT,BELLINI}: 
\beq 
\Gamma (H_Q) \simeq G_F^2 m_Q^5(\mu ) \cdot 
\left[ c_0(\mu ) + \frac{c_2(\mu )}{m_Q^2(\mu )}+ 
\frac{c_3(\mu )}{m_Q^3(\mu )}+ ... 
\right] 
\label{GAMMA}
\eeq
Some points have to be emphasized for later reference: 
\begin{itemize}
\item 
With the leading contribution to the width being 
proportional to the fifth power of the heavy quark mass it is 
essential to give an unambiguous and precise definition of the 
latter. This requires a field theoretical treatment 
specifying both the
renormalization scheme employed and the scale 
$\mu$ at which the mass is taken 
\footnote{An observable of course cannot depend 
on this scale $\mu$; the dependance of $m_Q$ has to be 
compensated for by a $\mu$ dependance of the radiative corrections,  
matrix elements etc., as denoted by the coefficients 
$c_i$ in Eq.(\ref{GAMMA}).}. Such a treatment has 
been given in terms of the so-called kinetic mass \cite{HQT} 
and the large sensitivity to the quark mass does not pose a 
real problem for the OPE.  
A quark model treatment
without OPE support  {\em cannot} meet this essential requirement as a
matter  of principle; it thus
suffers  from {\em intrinsic} ambiguities in the quark mass values, 
which lead to considerable irreducible uncertainties in its 
estimates for the weak widths.
This is  well-known and has been discussed extensively in the 
literature. 
\item 
The colour gauge symmetry forbids 
contributions of order $1/m_Q$. Hence: 
\begin{itemize}
\item 
Through ${\cal O}(1/m_Q)$ the contributions to the width are 
universal for a given flavour; i.e., they do {\em not} depend on the 
specific hadron. 
\item 
In particular, an ansatz where the leading term in the 
width is proportional to the fifth power of the heavy {\em hadron} mass 
rather than the heavy {\em quark} mass is inconsistent with the OPE. 
A clear-cut violation of quark-hadron  duality (duality) would be 
required for such an ansatz to emerge. Yet even then it could
hold  only for some values of the hadron mass, but not parametrically 
\cite{VADE}.  
\item 
The pole mass, though universal for a given flavour, cannot be used here
in principle, since it suffers from  an irreducible uncertainty $\sim
{\cal O}(\bar\Lambda )$; it would introduce an uncertainty 
$\sim 5 \bar \Lambda /m_Q$ into $\Gamma (H_Q)$, which is 
of {\em higher} order than the contributions one can calculate. 
\end{itemize}
\item 
The contributions of order $1/m_Q^2$ differentiate between 
baryons and mesons, while treating the latter in a practically 
universal way. 
\item 
In order $1/m_Q^3$ explicitely spectator-dependant contributions 
enter that are usually referred to as Pauli interference ($PI$) 
\cite{BRANCO}, weak 
annihilation ($WA$) and $W$ scattering. Those had been identified in 
quark model treatments already before the development of the 
HQE. 
\item 
Spectator and (higher order) $WA$ amplitudes can interfere 
with each other! This property is actually 
essential for the decay width to be controlled by the large 
scale $m_Q$, see Eq.(\ref{GAMMA}), rather than a hadronic low energy
scale, as has 
been demonstrated explicitely in Ref. \cite{MIRAGE}. As 
discussed in detail in Refs. \cite{WA} it is quite possible a priori  
that the inclusion of $WA$, when it is not the leading contribution, 
can actually diminish a width!    
\item 
With the OPE-based HQE providing a quantitative description of 
nonperturbative dynamics in 
heavy flavour decays, we have now reached a stage where 
we can go beyond folklore when discussing duality 
\cite{VADE}.   

\end{itemize}
 
\section{Criticism of Ref. \cite{NUSS}}
After this review in a nut shell I list specific criticism of various
statements in  Ref.\cite{NUSS}; the first point is basically 
conceptual, the others somewhat more technical: 
\begin{enumerate}
\item 
The authors suggest that the observed $D^0 - D_s^+$ lifetime 
difference is due to  
$SU(3)_{Fl}$ breaking in
the phase space of the exclusive $D^0$ and $D_s^+$ 
channels. For their argument they adopt the following prescription: they
take  all observed $D^0$ modes, which actually saturate the $D^0$ 
width, correct them one by one with some phase space factor for the 
corresponding $D_s$ modes and add up the thus obtained estimates 
for the partial
$D_s$ widths; this procedure yields $\tau (D_s^+)/\tau (D^0) 
\sim 1.25$ -- similar indeed to the observed value.  
\begin{itemize}
\item 
Such a description cannot be viewed as a genuine explanation, though.
Adding   all exclusive modes has to yield the total width, of course; the 
only nontrivial element here is that one relates $D^0$ and 
$D_s^+$ channels 
by simple phase space considerations. It has to be 
kept in mind, however, that there is no unique prescription 
for the phase space of multibody final states, since it depends very 
much on their resonance etc. structure. 
\item 
The real challenge is whether one can provide an at least 
adequate theoretical description of the 
hadron's width expressed on the 
quark-gluon level; that is the very essence of 
the concept of duality! A success in
this task signals that  duality applies already at the charm scale for
inclusive  transitions. It is on the quark-gluon level that 
$WA$ is 
primarily defined. 
\item 
In properly addressing duality one has to keep some subtle 
features in mind. Observable rates depend on the phase space as 
it applies to hadrons and their masses; it is certainly different from 
the phase space for quarks (and gluons). 
Duality cannot be based on some kind of equivalence between 
hadronic and quark-level phase space alone; it can hold only if 
quark phase space {\em plus} quark dynamics are effectively 
equivalent to hadronic phase space {\em plus} boundstate 
dynamics. Such equivalence, which at first sight would look 
miraculous, is enforced through sum rules -- 
as discussed in detail in Ref.\cite{FIVE}. 
 
An effect that on the {\em hadronic} level represents 
$SU(3)_{Fl}$ 
breaking merely in the phase space, has on the {\em quark} 
level  
to be a combination of $SU(3)_{Fl}$ breaking in the phase
space  {\em and} in quark dynamics. 
$SU(3)_{Fl}$ breaking in the quark phase space is
modest since 
$m_c \gg \Lambda \gg m_s, m_d, m_u$ (for {\em current} 
quark masses); likewise for the nonperturbative contributions 
$\sim {\cal O}(1/m_Q^2)$. In order $1/m_c^3$ $PI$ and 
$WA$ enter, both of which induce $SU(3)_{Fl}$ breaking in a 
quark level description. 
\end{itemize} 
\item 
Replacing the quark mass by an {\em effective} quark mass reduced by 
some sort of binding energy $BE$ is  
inconsistent with the OPE! For substituting $m_Q - BE$ for $m_Q$ 
in $\Gamma \propto m_Q^5$ generates a correction of order 
$5\cdot BE/m_Q$ which is {\em not} allowed due to the absence of 
an independant colour gauge invariant operator of dimension four.  
The absence of a ${\cal O}(1/m_Q)$ contribution can be 
understood in a less abstract way as well: there are indeed 
${\cal O}(1/m_Q)$ corrections to the initial state energy 
and likewise to the final state energies. Both of these 
effects yield ${\cal O}(1/m_Q)$ contributions to the phase 
space -- yet they compensate each other there up to terms  
$\sim {\cal O}(1/m_Q^2)$. This compensation is due to colour 
symmetry, namely that both initial and final state quarks 
carry the same colour charge.   
 
Replacing the quark mass by an effective mass reduced by some 
binding energy had been
suggested before in the description  of $B_c$ decays where it would have
some dramatic consequences:  it would shift the bulk of the $B_c$ width
to $b$ rather  than $c$ quark decays and lead to a relatively long $B_c$
lifetime  in excess of 1 psec. Yet it was pointed out 
\cite{BC,BEN} that the 
HQE derived from the OPE, 
which does not contain $1/m_Q$ terms, predicts the $B_c$ width to be
generated  mainly
by $c$ quark decays producing a `short' lifetime below  1 psec -- as has
subsequently been observed \cite{CDF}. 
 
The authors of Ref. \cite{ALTA} have proposed 
a different ansatz leading to $1/m_Q$ contributions, 
namely to replace the
heavy {\em quark} mass by the  heavy flavour {\em hadron} mass in the
expression for the total  width; it might be amusing to note in 
passing that such an ansatz would {\em reduce}  
$\tau (D_s)/\tau (D^0)$ by 30 \%.  
It is authors' privilege to suggest an ad-hoc ansatz; yet one should 
be clear on the price such an ansatz entails. In this case it would 
be a clear-cut breakdown in the OPE.  
Also the authors of Ref.\cite{NUSS} seem to be quite unaware that
questions of duality  and its limitations in observables described by an
OPE can be  addressed in a much more mature way now than five to ten
years ago 
\cite{VADE}.  
\item 
The first footnote of \cite{NUSS} claims that as far as 
inclusive transitions are concerned the amplitudes 
for the spectator and $WA$ processes have to be added in 
modulus squared  
with {\em no} interference since the final states correspond to 
four- and two-quark states, respectively. As already mentioned, 
such a statement is incorrect as demonstrated in the literature  
\cite{MIRAGE,WA}: the numerically leading $WA$ 
contributions  are provided by non-factorizable terms that are not 
helicity suppressed;  those are of higher
order in $\alpha _S$ and interfere with the 
dominant spectator amplitude 
(and others in general). A priori, this interference can  
be destructive as well as constructive. 
\item 
One can{\em not} argue that the $WA$ contribution is much more 
reduced in $D_s$ than in $D^0$ decays. While this would be true 
for the factorizable terms, it is not true in general for the 
non-factorizable ones which are numerically more 
important due to the chiral suppression of the factorizable 
terms \cite{WA}.  
 
\item 
There are some more detailed tests of the HQE description of 
$D_s$ decays. The width of Cabibbo suppressed $D_s$ decays 
leading to final states with the quantum numbers of 
$u \bar s s \bar s$ will be reduced by a factor of about two 
due to $PI$, whereas the latter cannot affect the 
corresponding $D^0$ transitions; likewise for doubly 
Cabibbo suppressed modes. This relative suppression is in 
both cases considerably larger than in the ansatz of Ref. 
\cite{NUSS}.   
However, identifying the footprints of WA in subclasses of 
decays is still an open challenge.  

\end{enumerate}   

\section{Summary}
There is nothing wrong with coming up with a simple 
(or for that matter
even a sophisticated)  phase space based description for relating
exclusive
$D^0$  and $D_s$ channels leading to an estimate of 
$\tau (D_s)/\tau (D^0)$. Yet that does not remove the motivation  
for analysing whether one can find a `dual' quark based 
description; furthermore $WA$ is an element of non-perturbative 
quark rather than hadronic dynamics.  
Duality could not be realized by simply equating quark phase space 
with an average over hadronic phase space: it requires 
quark phase space {\em coupled} with quark dynamics to account for 
hadronic phase space together with boundstate effects on average.  
The OPE naturally incorporates quark decay, $PI$, $WA$ etc. -- 
and interferences between them -- in a well-defined way. It is
certainly conceivable that the OPE based description fails for 
charm decays since the charm scale could be too marginal for 
duality to apply here. In that case -- but only in that case -- 
contributions of order $1/m_Q$, as would be introduced through 
use of an effective quark mass $m_Q - BE$, could emerge. 
But it should be clearly understood and explicitely stated that 
such an occurrance would constitute a clearcut breakdown of 
the OPE!   
\vspace*{.2cm}  
{\bf Acknowledgements:}~~
This work has been supported by the 
National Science 
Foundation under grant number PHY00-87419. 
 

\end{document}